# Teaching Experiences using the RVfpga Package

D. Chaver, S. Harris, L. Piñuel, O. Kindgren, R. Kravitz, J. I. Gómez, F. Castro, K. Olcoz, J. Villalba, A. Grinshpun, F. Gabbay, L. Seed, R. Duarte, M. López, O. Alonso and R. Owen

***Abstract*—RISC-V is an open-standard instruction set architecture (ISA) based on reduced instruction set computing (RISC) principles. Unlike proprietary ISAs, RISC-V is free and open, allowing anyone to design, manufacture, and sell RISC-V chips and software. Its simplicity and modular design make it highly customizable, catering to a wide range of applications, from small embedded systems to powerful supercomputers. RISC-V's flexibility, combined with its growing ecosystem and community support, has made it increasingly popular for research, education, and commercial development in the semiconductor industry. The RISC-V FPGA (RVfpga) course offers a comprehensive introduction to computer architecture using the RISC-V instruction set and FPGA technology. It focuses on providing hands-on experience with real-world RISC-V cores, the VeeR EH1 and the VeeR EL2, developed by Western Digital a few years ago and currently hosted by ChipsAlliance. This course is particularly aimed at educators and students in computer science, computer engineering, and related fields, enabling them to integrate practical RISC-V knowledge into their curricula. The course materials, which include detailed labs and setup guides, are available for free through the Imagination University Programme website. We have used RVfpga in different teaching activities and we plan to continue using it in the future. Specifically, we have used RVfpga as the main experimental platform in several degree/master courses; we have completed several final degree/master projects based on this platform; we will conduct a microcredential about processor design based on RVfpga; we have adapted RVfpga to a MOOC in the edX platform; and we have shared RVfpga worldwide through one-day hands-on workshops and tutorials. This paper begins by discussing how the RVfpga course matches the latest IEEE/ACM/AAAI computing curriculum guidelines. It then details various teaching implementations we have conducted over recent years using these materials. Finally, the paper examines other courses similar to RVfpga, comparing their strengths and weaknesses.**

## I. INTRODUCTION

RISC-V is a free and open-source instruction set architecture (ISA) introduced in 2010 at the University of California, Berkeley. Its design adheres to the principles of the reduced instruction set computer (RISC) philosophy, emphasizing simplicity and efficiency. Since its inception, RISC-V has gained widespread adoption across industry, academia, and research, thanks to its flexibility, scalability, modularity and absence of licensing fees, making it an attractive choice for a wide range of computing applications. The RISC-V architects established five main principles for its design:

- Compatibility: It must be compatible with a wide range of software packages and programming languages.
- Feasibility: Its implementation must be feasible across various technology options, including FPGAs (field-programmable gate arrays), ASICs (application-specific integrated circuits), and emerging technologies.
- Efficiency: It must be efficient across various microarchitectural scenarios, including those using microcode or hardwired control, in-order or out-of-order pipelines, and various types of parallelism.
- Customization: It must be adaptable to specific tasks to achieve optimal performance without limitations imposed by the ISA itself.
- Stability: Its base instruction set must be stable and long-lasting, providing a common and robust framework for developers.

The ecosystem built around the RISC-V architecture, including the toolchain, processors and simulators, has been steadily developing since RISC-V's inception in 2010. Despite these developments, the ability to readily access, use, and work with the RISC-V architecture and open-source RISC-V systems still faces barriers, including lack of understanding the architecture and its extensions, the inability to access, understand, and use RISC-V tools, and the difficulty in gathering and using all of the pieces needed to deploy, understand, and expand a commercial, non-trivial, RISC-V core and SoC with its accompanying development environment and tools. While some courses address some of these barriers, the RVfpga courses (the **RVfpga in Computer Architecture** [1] and the **RVfpga-SoC in SoC design** [2]) aim to address all of these barriers by showing the complete path from system setup to using, programming, understanding, and expanding a RISC-V core and SoC.

In RVfpga we provide two SoCs, one based on the **VeeR EH1** core [3] (see ***Fig. 1***) and one based on the **VeeR EL2** core [4] (the SoC is the same as the one shown in ***Fig. 1***, except that it uses the smaller EL2 core). The EH1-based SoC includes the following peripherals: Boot ROM, System Controller, two SPIs, PTC, GPIO and UART. The SoC uses a 64-bit AXI bus to communicate with the VeeR EH1 core, and a Wishbone bus to connect with the peripherals. In addition to providing the RVfpga SoC source code in Verilog/SystemVerilog, RVfpga shows how to install and use the RISC-V toolchain to compile, debug, and run C and RISC-V assembly programs onto the SoC, and how to use the peripherals provided with the SoC and expand the SoC to add new peripherals.

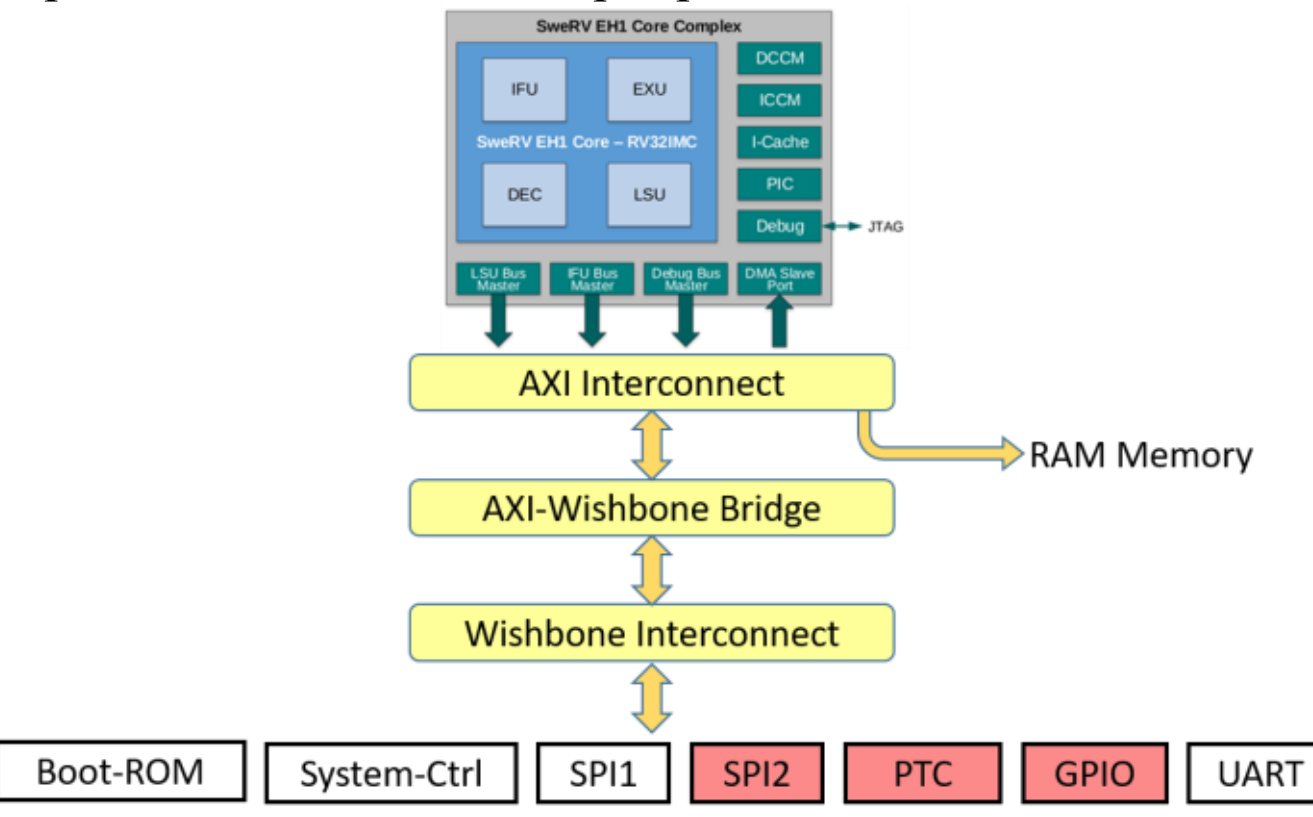

***Fig. 1*** The VeeR EH1 based RVfpga SoC

The RVfpga materials also show how to explore and modify the VeeR EH1/EL2 core's microarchitecture, including adding instructions to the core, measuring performance using built-in performance counters, and exploring microarchitectural features, from the most fundamental aspects, such as pipelining, caches, and hazards, to other more advanced capabilities, such as superscalar execution and scratchpad memories. TABLE I summarizes the 20 labs included in the package.

TABLE I
RVFPGA LABS

| Lab 1 | C Programming |
|---|---|
| Lab 2 | RISC-V Assembly Language |
| Lab 3 | Function Calls |
| Lab 4 | Image Processing: Projects with C & Assembly |
| Lab 5 | Creating a Vivado Project |
| Lab 6 | Introduction to I/O |
| Lab 7 | 7-Segment Displays |
| Lab 8 | Timers |
| Lab 9 | Interrupt-Driven I/O |
| Lab 10 | Serial Buses |
| Lab 11 | VeeR Configuration and Organization. Performance Counters |
| Lab 12 | Arithmetic/Logical Instructions: `add` |
| Lab 13 | Memory Instructions: the `lw` and `sw` Instructions |
| Lab 14 | Structural Hazards |
| Lab 15 | Data Hazards |
| Lab 16 | Control Hazards: Branch Instructions. The Branch Predictor |
| Lab 17 | Superscalar Execution (only for VeeR EH1) |
| Lab 18 | Adding New Instructions to the Core |
| Lab 19 | Memory Hierarchy: The Instruction Cache (I$) |
| Lab 20 | I$, ICCM, DCCM, and Benchmarking |

The second course, RVfpga-SoC [2], is a short follow-on course to RVfpga that shows how to build a RISC-V SoC from scratch using building blocks and then how to run the Zephyr real-time operating system (RTOS) on it. TABLE *II* summarizes the 5 labs included in the package.

TABLE II
RVFPGA-SOC LABS

| Lab 1 | Introduction to RVfpga-SoC |
|---|---|
| Lab 2 | Running Software on the RVfpga SoC |
| Lab 3 | Introduction to SweRVolf and FuseSoC |
| Lab 4 | Building and Running Zephyr on the RVfpga SoC |
| Lab 5 | Running Tensorflow Lite on SweRVolf |

The RVfpga SoC can be run on different FPGA boards, from low-cost ones such as **Basys3** [5] or **Boolean** [6] boards to more expensive ones such as the **Nexys A7** board [7]. The RVfpga SoC can also run on different simulation tools: **RVfpga-ViDBo**, that simulates the RVfpga SoC running a program using a web-based virtual board to interact with the FPGA peripherals (see ***Fig. 2***); **RVfpga-Pipeline**, that shows the instructions executing through the VeeR EH1/EL2 pipeline as it simulates a program running on the RVfpga SoC; **RVfpga-Trace**, that shows a waveform of the system's internal signals during execution of a program running on the RVfpga SoC; and **Whisper**, a RISC-V instruction-set simulator (ISS).

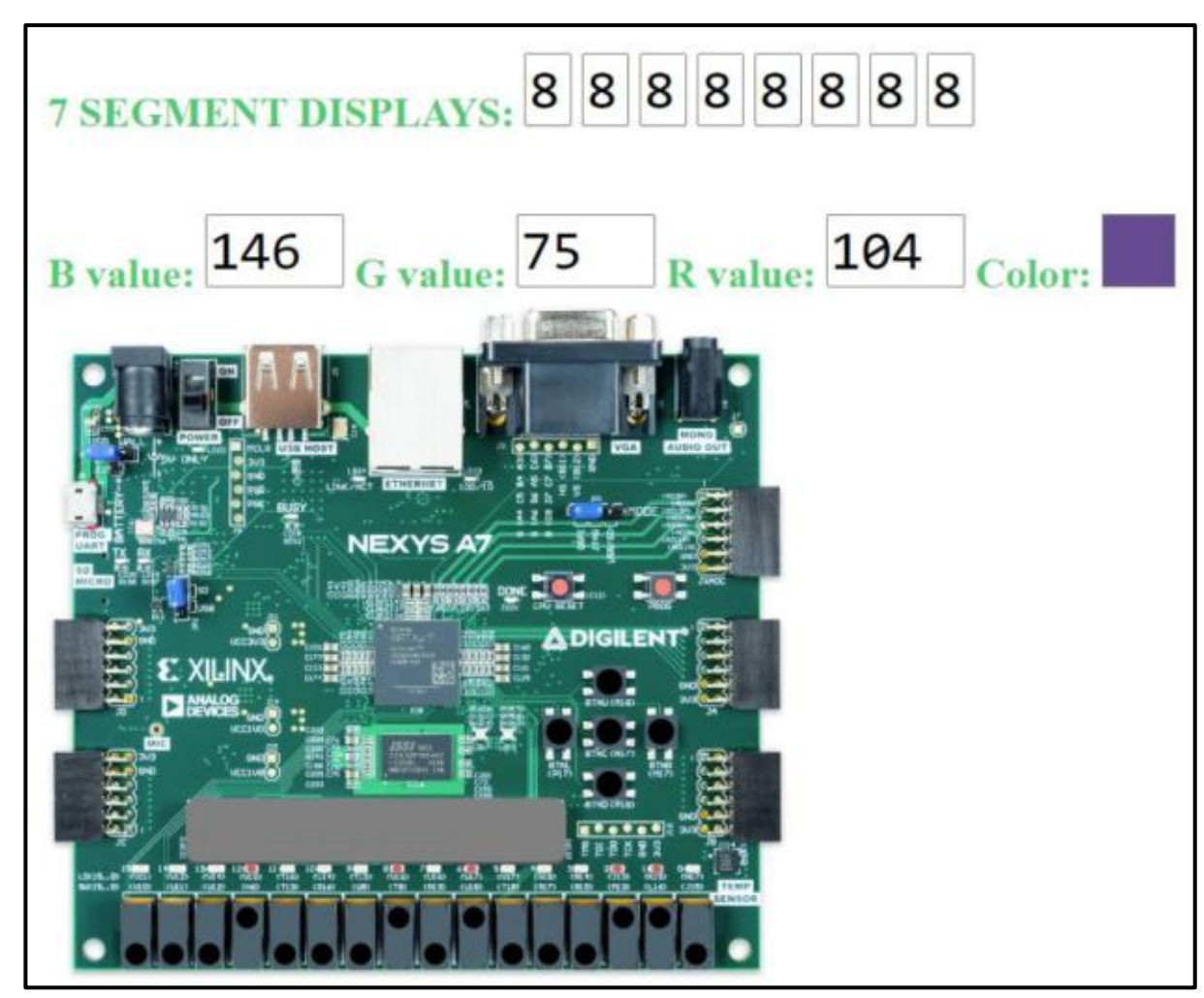

***Fig. 2*** An example of the RVfpga-ViDBo running a program that communicates with the 16 switches, the 16 LEDs, and the tri-color LED on the Nexys A7 FPGA board

We recently published several papers [9] [10], where the RVfpga courses are described in detail. In this new paper, we first describe how the RVfpga course meets the most recent IEEE/ACM/AAAI computing curriculum guidelines. We then describe various teaching applications that we've carried out in recent years based on these materials, including the adaptation to a MOOC in the edX platform [11]. Finally, we describe other available courses similar to RVfpga, highlighting their strengths and weaknesses.

## II. COMPUTING CURRICULUM GUIDELINES IN RVFPGA

RVfpga covers the topics in the most recent computing curriculum guidelines developed by relevant associations such as ACM or IEEE in the Architecture and Organization (AR) area.

Several successive curricular guidelines for computing studies have been published over the years as the discipline of Architecture and Organization has continued to evolve. The Computer Science Curricula 2023 Final Report [12], also referred to as CS2023, is the latest version of these types of guidelines, produced by a joint task force of the ACM, IEEE Computer Society, and AAAI. Another older but wider report can be found at [13], which, in addition to Computer Science, analyzes other computing areas such as Computer Engineering or Software Engineering.

The CS2023 knowledge model consists of 17 knowledge areas. Among these, the one that RVfpga seeks to cover is the Architecture and Organization (AR) area, described in detail in a dedicated section of the report (pages 101-111 of [12]). That section states in its preamble that, as the shift from sequential to parallel processing occurs, a deeper understanding of the underlying architecture is necessary, which can no longer be viewed as a black box where principles from one architecture can be applied to another one. Instead, programmers should look inside the black box and use specific components to enhance system performance and energy efficiency. The AR knowledge area aims to develop a deeper understanding of the hardware environments upon which almost all computing is based, and the relevant interfaces provided to higher software layers. The topics in this knowledge area will benefit students by enabling them to appreciate the fundamental architectural principles of modern computer systems, including the challenge of harnessing parallelism to sustain performance and energy improvements into the future, and will help students depart from the black box approach and become more aware of the underlying computer system and the efficiencies specific architectures can achieve.

Specifically, the following knowledge units are established in the AR area, and they are fully or partially covered in RVfpga:

- Digital Logic and Digital Systems: This unit includes topics such as FPGAs, Hardware Description Languages (HDLs) and SoC design flow, which are exhaustively covered in the RVfpga packages. For example, the RVfpga-SoC course covers SoC design flow, whereas labs 6 to 10 of RVfpga provide exercises where the student must understand and modify the I/O controllers, which are coded in Verilog/SystemVerilog, and run various programs on the RVfpga SoC on an FPGA board.
- Machine-Level Data Representation: This unit includes topics such as numeric data representations and number bases such as fixed-point and floating-point, which are discussed in RVfpga. For example, Lab 18 includes a comprehensive exercise where the student must add a floating point unit (FPU) to the VeeR core and then program, execute and compare the performance of different floating-point algorithms.
- Assembly Level Machine Organization: Topics such as the instruction set architecture (ISA) (e.g., x86, ARM and RISC-V) and subroutine call and return mechanisms are included in this unit, and these topics are exhaustively covered in RVfpga. For example, RVfpga labs 2 to 4 describe the RISC-V ISA and provide numerous examples and exercises for the students to practice these concepts.
- Memory Hierarchy: This unit includes topics such as main memory organization and cache memories, which are covered in RVfpga. For example, lab 19 describes the Instruction Cache used in the VeeR cores.
- Interfacing and Communication: I/O fundamentals (such as programmed I/O vs. interrupt-driven I/O), interrupt structures (vectored and prioritized), I/O devices and fundamentals of buses are included in this unit, which are covered in RVfpga. For example, RVfpga lab 9 focuses on the use of interrupts in the RVfpga SoC and the differences between programmed I/O vs. interrupt-driven I/O.
- Functional Organization: This unit includes topics such as implementation of simple datapaths, instruction pipelining and hazard detection, which are exhaustively covered in RVfpga. For example, labs 11-13 analyze pipelined execution in the VeeR core and labs 14-16 analyze the different types of hazards: structural, data and control.
- Performance and Energy Efficiency: This unit includes topics such as performance, power consumption, memory evaluation, branch prediction, speculative execution, hardware support for multithreading, etc. Some of the topics covered in this unit are included in RVfpga; for example, Lab 20 performs performance evaluation using performance counters and two well-known benchmarks (Coremark and Dhrystone), and Lab 16 analyzes branch prediction.

Note that the remaining units in this knowledge area (Heterogeneous Architectures, Secure Processor Architectures, and Quantum Architectures) are not covered in the current RVfpga packages, but some of them could be included by extending the packages. For example, we could add support for VeeR EH2, a dual-threaded core, which would allow us to include labs that analyze the hardware support for multithreading topics mentioned in the Heterogeneous Architectures unit. We could also add a lab using the RISC-V bit manipulation extension, implemented in VeeR, in cryptographic applications, thus covering the Cryptographic Acceleration with Hardware topic mentioned in the Secure Processor Architectures unit.

As for the other report mentioned above [13], it refers not only to Computer Science but to six more computing areas: Computer Engineering, Cybersecurity, Information Systems, Information Technology, Software Engineering, and Data Science. This report defines thirty-four abbreviated knowledge areas for computing students partitioned into an ordered sequence of six categories, one of which is devoted to Hardware, which includes knowledge areas covered in RVfpga, such as Architecture and Organization and Digital Design; a second category is devoted to Systems Architecture and Infrastructure, which includes knowledge areas also covered in RVfpga, such as embedded systems. Example topics included in these knowledge areas are control and datapaths, programmable logic, processor organization, memory system organization and architecture, and I/O interfacing, all of which are exhaustively analyzed and discussed in RVfpga.

## III. TEACHING EXPERIENCES USING RVFPGA

In this section, several professors from different universities around the world describe their experience using RVfpga in their formal degree/master teaching activities (sections A and B). Then, we describe other teaching-like experiences related with RVfpga (sections C, D and E), including workshops, MOOCs, and research.

### *A. Degree and Master courses*

Several universities are using RVfpga in their Computer Architecture courses. In this section we describe some of these experiences; specifically:

1. Integrated Systems Architecture at University Complutense of Madrid (UCM)
2. Computer Organization at UCM
3. SoC Design with Programmable Logic at Portland State University
4. RISC-V Hackathon at the Ruppin Academic Center
5. Introduction to Computers and Computers Architecture at UB

We are also developing a masters program focusing on processor design, where we will include some labs based on RVfpga.

#### A.1. Integrated Systems Architecture at UCM

This is a fourth-year course in the Electronics and Communication Engineering degree that we offer at UCM. The students of this course are in the last year of their degree and have abackground in digital design, hardware description languages (HDLs), computer architecture/organization, C programming and operating systems; thus they have enough skills to complete most of the RVfpga and RVfpga-SoC labs, and only lack of time precludes us from including the whole packages. At [14] you can find all the slides, documents and labs used in the course.

The course starts with an introduction to the most recent trends in computer architecture, and then it explores advanced processor design (deep pipelining, multi-cycle operations, speculative execution and branch prediction), the memory system (memory hierarchy, scratchpad memories), the I/O system, benchmarking and SoC design.

The course includes 5 modules and 9 labs (see TABLE *III*), distributed across 30 1.5-hour lectures plus 15 weekly 2-hour lab sessions. In the labs, we use both the RVfpga and RVfpga-SoC packages, as well as the Ripes simulator [15], a visual computer architecture simulator and assembly code editor built for the RISC-V ISA. Each student is given a Nexys A7 FPGA boardfor the semester. The VeeR EH1 core fits on this board and it is a more advanced processor than EL2, which allows us to explore concepts such as deep pipelining and superscalar execution.

TABLE III
MODULES AND LABS INCLUDED IN THE INTEGRATED SYSTEMS ARCHITECTURE COURSE AT UCM

|  |  |
|---|---|
| **Module 1: RISC-V ISA** | Lab 1: RISC-V ISA review |
| **Module 2: Processor Design** | Lab 2: Introduction to the 5-stage core by Harris&Harris and to the VeeR EH1 core |
| | Lab 3: Performance of the 5-stage and VeeR EH1 cores |
| | Lab 4: Execution of basic instructions on the VeeR EH1 core |
| **Module 3: Memory Hierarchy** | Lab 5: The cache on the Ripes simulator and on the VeeR EH1 core |
| | Lab 6: The scratchpad on the VeeR EH1 core |
| **Module 4: Input/Output** | Lab 7: High-level I/O on the RVfpga SoC |
| | Lab 8: Low-level I/O on the RVfpga SoC |
| **Module 5: SoC Design** | Lab 9: RVfpga-SoC |

Before the first lab session, students install the provided virtual machine, which includes all the tools installed and ready to use. Although using the virtual machine is the most convenient approach, as it avoids installation problems or difficulties associated with the use of different operating systems, students may also install and use the software tools natively on their computers following the steps described in the Getting Started Guide.

Module 1 reviews the RISC-V ISA. This module's lab (Lab 1) describes how to execute programs on the Nexys A7 FPGA board (see ***Fig. 2***) and on the Whisper simulator (which can be used when a board is not available), and then the students complete RVfpga Lab 4 (see TABLE *I*). In Lab 4, an image transformation algorithm is provided, that transforms an RGB image into a greyscale image. The project includes multiple source files, some of which are written in C and some in assembly, and it shows how C functions can invoke assembly routines and vice versa. The students must understand and execute the provided project and then complete some exercises, where some of the C/assembly functions must be transformed into equivalent assembly/C functions and where different filters (such as a Blur Filter) must be programmed and tested on different images.

In Module 2 we review the 5-stage pipelined processor provided in [16], which the students already know from previous courses. Then, we explain the microarchitecture of the VeeR EH1 processor, which is new to the students. We devote several lectures to understanding the microarchitecture of the VeeR EH1 processor, its deep 9-stage pipeline, and its five datapaths (I0/I1: dual integer execution, LS: load/store, Mult: multiplication and Div: divide), its superscalar execution, the multi-cycle operations that it supports (loads/stores and multiplication), its forwarding paths, its memory hierarchy, and its configurability in both software and hardware. The module contains three labs: Labs 2-4.

In Lab 2, we show students how to use the Ripes and RVfpga-Pipeline simulators (***Fig. 3*** illustrates an example of the RVfpga-Pipeline simulator running a program; you can find more detailed descriptions for the simulator in [14]) and we propose exercises where simple RISC-V assembly programs are provided and must be analyzed first theoretically and then experimentally using the simulators. The programs illustrate situations such as structural and data hazards, branch instructions and control hazards using different branch predictors, the use of the Secondary ALU, etc.

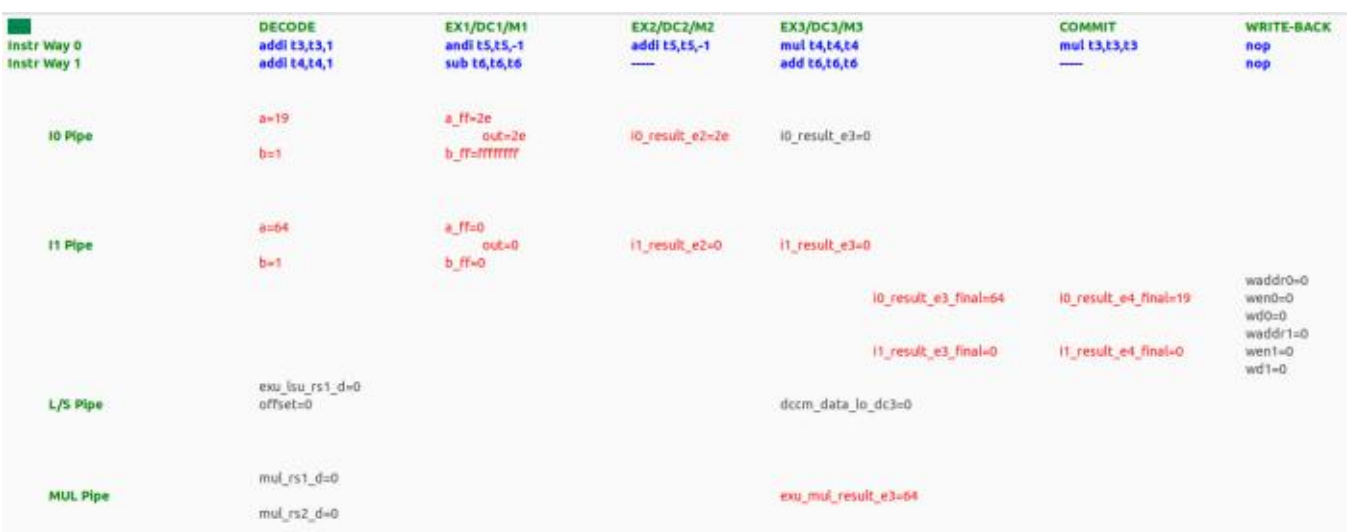

***Fig. 3*** An example of the RVfpga-Pipeline simulator running a program

In Lab 3, we show how to use the RVfpga-ViDBo simulator [14] (***Fig. 2***), which can be used in place of the Nexys A7 board, and we explain how the VeeR core can be configured. Then, we propose exercises where performance must be measured for various RISC-V assembly and C programs running on different core configurations. To measure performance, we explain how to use Western Digital's Processor Support Package (PSP) to configure and use the performance counters in the VeeR core.

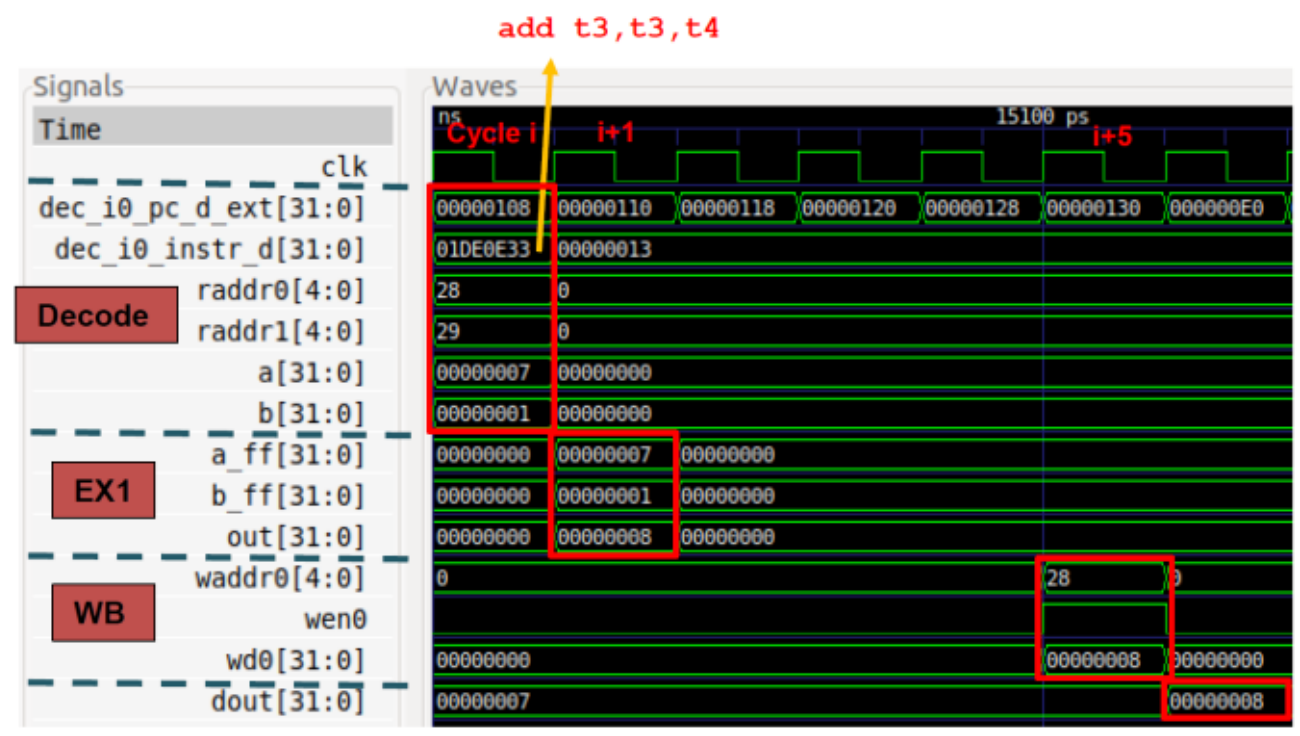

***Fig. 4*** An example of a waveform generated by the RVfpga-Trace simulator for the execution of an `add` instruction

Finally, in Lab 4, we show how to use the RVfpga-Trace simulator (***Fig. 4*** illustrates an example of a waveform generated by this simulator), we provide a condensed version of RVfpga Lab 12 (see TABLE *I*), where students must analyze the execution of an `add` instruction on the VeeR EH1 pipeline. Several exercises are also proposed, where students must analyze other RISC-V Arithmetic-Logic instructions.

In Module 3, we first review memory hierarchy fundamentals (cache, main memory, virtual memory, scratchpads, etc.), using [16], and then we propose two labs: Lab 5 and 6. Lab 5, which uses the Ripes and RVfpga simulators. The RVfpga SoC does not include a data cache (D$), thus we study a D$ in Ripes doing simple exercises. Then, given that the RVfpga SoC does include an instruction cache (I$), we show more advanced exercises based on Lab 19 of RVfpga (see TABLE *I*). Lab 6 is based on RVfpga Lab 20 (see TABLE *I*), which uses and analyzes scratchpad memories (ICCM and DCCM) available in the VeeR EH1 processor. Two commercial benchmarks (Coremark and Dhrystone) are used to compare different configurations of the VeeR EH1 core and see how the scratchpad can significantly improve performance under some circumstances.

In Module 4, we explain in a few lectures the main I/O concepts as well as the RVfpga I/O peripheral system shown in Fig.1: the I/O memory map, the interrupt system, SPI, etc. Then the students complete the guided examples and selected exercises across several lab sessions. Lab 7 includes some basic exercises from Labs 6-10 of RVfpga (see TABLE *I*), including developing programs to communicate with the peripherals provided in the baseline RVfpga SoC. Lab 8 includes some advanced exercises from Labs 6-10 of RVfpga (see TABLE *I*), where students must extend the I/O system with new peripherals by extending the SoC using SystemVerilog; specifically, they have to add pushbuttons and extend the functionality of the 7-segment displays. The students then develop programs to exercise these added peripherals.

Finally, in Module 5, using the RVfpga-SoC package, the students build the RVfpga SoC from scratch using FuseSoC [17] and build, run, and use Zephyr on the new RVfpga SoC.

The students that attended the lab sessions were able to complete all or most of the labs. According to the marks obtained and to their opinions in the final survey, the labs were a valuable resource for complementing and reinforcing the theoretical concepts.

### A.2. Computer Organization at UCM

This is a second-year course in the Computer Science degree program offered at UCM. Traditionally, this has been an entirely theoretical subject at UCM, leaving students without the opportunity to use or apply the concepts in practice. However, as with many fields, in the field of computing, many concepts can only be understood when they are both studied theoretically and practiced experimentally. Therefore, last year we decided to include 10 hours of lab sessions, based on RVfpga, to complement and reinforce the theoretical concepts. In [14] you can find all the slides, documents and labs used on the course.

Compared to the Integrated Systems Architecture course discussed earlier, students in this course have less background, less time for labs, and the enrollment is higher. At this point, they have only studied Computer Fundamentals, where they only begin to learn the fundamentals of Digital Design and RISC-V Computer Architecture, so the exercises selected cannot be too advanced. Also, because enrollment is higher and we do not have enough FPGA boards for all students, the exercises must be completed in simulation only (specifically, using Whisper, RVfpga-ViDBo, Ripes and RVfpga-Pipeline).

The theoretical aspect of the course starts with a detailed review of those concepts introduced in their first year: RISC-V

architecture and assembly programming, and the single/multi-cycle and pipelined processors from [16]. Then, the VeeR EH1 pipelined processor is used to introduce some advanced microarchitectural techniques, such as deep pipelining, superscalar execution, and multicycle operations. Finally, a generic I/O system and memory system are analyzed in detail.

The course includes a subset of the labs, in some cases simplified, shown in TABLE *III* for the Integrated Systems Architecture course discussed earlier. For the labs, the students work in groups of four, again using the Virtual Machine installed on their own laptops.

The first lab is a review of the RISC-V architecture and is analogous to Lab 1 in TABLE *III*. However, this time everything must be executed in Whisper, and the exercises are simpler, mainly selected exercises from Lab 2 of RVfpga (see TABLE *I*). Lab 2 is the same as Lab 2 in TABLE *III*; it uses the Ripes and RVfpga-Pipeline simulators (see ***Fig. 3***) to analyze microarchitectural techniques. Lab 3 is a simplified version of Lab 5 in TABLE *III*; it analyzes the cache using the Ripes simulator. Finally, Lab 4 is the same as Lab 7 in TABLE *III*, with the only difference that the exercises must be executed on RVfpga-ViDBo (***Fig. 2***); this lab gives hands-on exercises about I/O on RVfpga. The students do not extend RVfpga's I/O system, because they do not yet have knowledge of HDL programming.

After completing each lab, each student group must send a report that shows their exercise solutions accompanied by detailed explanations.

The results were quite good, as most students were able to successfully complete all labs and the final grades where higher than in previous years. In addition, at the end of the semester, the students must complete a survey about the course and they rated it very highly in their responses. Two student comments are:

- "The way the professor used the labs to clarify theoretical concepts should stand out as a model for other courses."
- "Personally, I'm not particularly passionate about Computer Architecture related subjects, I'm more into software. However, I must admit that this professor makes the subject enjoyable and the lab sessions clearly demonstrate the usefulness of the theoretical classes."

A.3. SoC Design with Programmable Logic at Portland State University

This course is part of a two-course sequence in the Masters in Electrical Engineering (MSEE) Program, Embedded Systems track at Portland State (PSU). The SoC Design course is taken at the end of the first or beginning of the second year, and it emphasizes the hardware aspects of implementing SoCs on an FPGA. The course is RVfpga-based and makes use of Imagination Technologies' Catapult IDE for software development. The target FPGA platforms are Digilent Nexys A7 and the RealDigital Boolean board. The Boolean board has proven to be a cost-effective FPGA board for those who choose to purchase their own hardware instead of using the Nexys A7 boards in the teaching labs.

The course is project-based; there are several projects, which take two weeks each, and a team-based final project that typically lasts the final three or four weeks (see TABLE *IV*). There are no homework assignments other than the projects, with final grades being based on the projects and two exams. The TA for the courses provides support to the students by hosting office hours in the labs and spending additional time with students or groups who are having problems. Students typically enter the course with experience in C programming and digital system design in SystemVerilog, but some students have little experience in the RISC-V microarchitecture or analyzing and modifying complex HDL code. Students are graded based on functionality as well as code readability; if the TA or professor cannot understand the HDL or application code, the grade is reduced.

The SoC Design course starts with an introduction to FPGAs and synthesis. Later lecture topics include an overview of the AMD/Xilinx Series 7 FPGA architecture, how to design to meet timing constraints, synchronization, and crossing clock domains. We then overview the RISC-V ISA, programming in assembly language, and the AXI and Wishbone buses. We also do a walkthrough of the RVfpga source code using a straightforward example to traverse the module hierarchy. It is a lot of material to pack into 10 weeks (PSU operates on quarters), so we don't have too much time to discuss the RISC-V architecture and microarchitecture.

TABLE IV
PROJECTS AND RVFPGA RESOURCES

| PROJECT | RVfpga RESOURCES |
|---|---|
| Getting Started | • RVfpga GSG<br>• Lab 2: RISC-V Assembly Language<br>• Lab 3: Function calls |
| Adding pushbuttons and RGB LED support to RVfpga | • Lab 5: Creating a Vivado Project<br>• Lab 6: Introduction to I/O<br>• Lab 7: 7-segment Displays |
| Implementing a simple VGA controller for RVfpga | |
| Final Project | Several final project reports have referenced:<br>• Lab 8: Timer<br>• Lab 9: Interrupt-Driven I/O<br>• Lab 10: Serial Buses |

The first project, Getting Started, heavily leverages the RVfpga Getting Started Guide (GSG). We ask the students to read and implement the sections of the GSG on installing and using the tools. The project is largely a self-study with a single deliverable - a 3- to 5-page design report describing their experience and pointing out any problems found along the way. We ask the students to include screenshots, logs, and/or videos to demonstrate that they have done the work.

In the second project, Adding an RGB PWM Controller to RVfpga, the students add several peripherals to the RVfpga-EL2 SoC. The students expand the RVfpga SoC to include new peripherals using SystemVerilog and Vivado. Students build two applications, one in RISC-V assembly language and the second in either RISC-V assembly or C. This project has three hardware tasks:

- Use Vivado to create, synthesize, and implement the expanded SoC from the original RVfpga source code.

- Add a second general-purpose I/O (GPIO) to the Wishbone bus to read the pushbuttons.
- Design, implement, and integrate a custom peripheral to control the RGB LEDs in the RVfpga SoC.

To keep the scope of the project reasonable for a 2-week project, we provide working SystemVerilog code for a 3-channel PWM generator. Students define a register map and encapsulate the PWM generator into a custom Wishbone peripheral.

The two applications the students must write are:

- An application written in RISC-V ssembly language; it should displays the value of a 16-bit counter on the LEDs. The pushbuttons are used to adjust the speed of counter and set the count to the value of the switches.
- An application written in either C or Assembly. The project description gives a specification for the application but no starter code. The application must copy the switches to the LEDs and use the buttons to increase the PWM duty cycles for the three segments. Each button press/release increments the duty cycle of one of the segments until it wraps around. The duty cycles are displayed on the 7-segment display

The third project is to adding VGA functionality to RVfpga. In our 10-week quarter, there is little time after project 2 for another large project. Yet one of the goals of the final project is for it to be visually interesting because the project is demonstrated in class. VGA provides just such a platform. So, this project asks the students to add a very simple VGA controller and application to RVfpga. The students are provided with a dot timing generator module which outputs a pixel row and column. Students add a 25 MHz clock to RVfpga to drive the VGA controller and create an additional Wishbone bus peripheral which displays a bit-mapped character on the display. The application provides a starting pixel row and column for the character so that the character can be positioned on the screen. In fact, the second part of the application is to build a screen saver that moves one or more characters around the display. The logic in the VGA peripheral handles displaying the character. We take advantage of the VGA-to-HDMI converter module provided by RealDigital to interface to monitors that have an HDMI input. While relatively simple in scope, this project leads to richer VGA controllers in students' final projects.

The SoC Design course culminates in a team-based final project. Students self-select into teams of 2-4 students each and then propose a final project. The instructors review the proposal and often negotiate deliverables and milestones. A proposal has both required and stretch goals. As the project progresses, if it appears that the students are running behind, we work with the teams to revise the proposal and expected deliverables. Even so, it is not unusual to have one or more projects fall short.

The grading rubric for the project includes a technical presentation of the project for the class, a demonstration of completed vs. required goals, the quality of the written report, and the readability of the HDL and application code. Each project receives a degree-of-difficulty score which is used to balance the projects and level the playing field. We award what I call the "Wall of Fame" award to the project which, in our opinion, was executed the best; this is not necessarily the most complicated project, but the project that has a clean proposal, excellent technical presentation, a demonstration that met committed goals, and outstanding documentation. The reward is bragging rights and a few extra credit points, but it is fair to say that many of the local hiring managers know about the "Wall of Fame" and what it represents.

For example, the "Wall of Fame" winner from winter term 2023 was a very playable Tetris game. The buttons on the NexysA7 board controlled the orientation and rate of fall for the pieces. The shapes and orientations of the pieces were stored in block RAM. The VGA controller from project 3 was redesigned for the application, and the software application controlled the game play. The project was a "Wall of Fame" winner because the team cleanly implemented what they proposed, the technical report was concise and added to my understanding of their application, and the code was organized well and documented – in fact, it was fun to read.

A.4. RISC-V Hackathon at the Ruppin Academic Center

The RISC-V hackathon has been introduced as a new course for undergraduate students in the computer engineering and electrical engineering departments at the Ruppin Academic Center. This hackathon challenge incorporates multidisciplinary, system-level technical skills, including microarchitecture, chip design, software development, and machine learning algorithms. The VeeR EL2-based RVfpga SoC platform was used in the hackathon and implemented on the Nexys A7 FPGA board. The Vivado Design Suite was used for hardware synthesis, analysis, simulation, debugging, and programming. Segger Embedded Studio served as the software development platform, providing an integrated environment with an editor, compiler, linker, simulator, and debugger. The overall hackathon kit, comprising the Nexys A7 FPGA, the Vivado Design Suite, Segger Embedded Studio, and a laptop PC, is illustrated in ***Fig. 5***. The PC was connected to the FPGA board using the designated USB cable and a J-Link EDU Mini adapter, allowing Segger Embedded Studio to run the VeeR EL2 core in debug mode via its JTAG port.

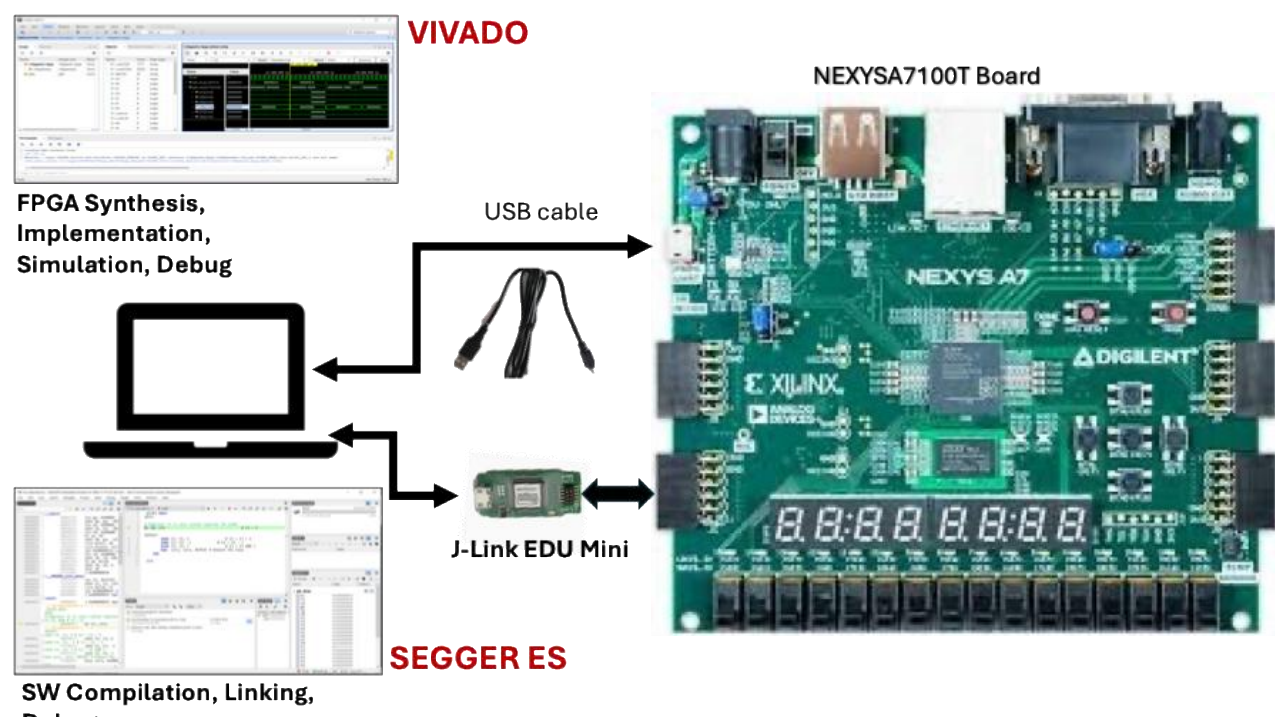

***Fig. 5*** The hackathon development platform

The hackathon kits were distributed to all participants, accompanied by two preparation workshops that took place two months before the hackathon event. The first workshop provided general training on the VeeR EL2-based RVfpga SoC platform, the Nexys A7 FPGA, the Vivado Design Suite, and Segger Embedded Studio. All students completed the full installation process of the hackathon platform, including the compilation and burn-in of the VeeR EL2 SoC, as well as compiling, loading, and running new programs written in C on the RISC-V SoC. Special attention was given in the workshop to the debugging capabilities of the hackathon platform, which included: 1) the Segger Embedded Studio debugger, allowing programs to run step-by-step, use breakpoints, and show full visibility of architectural registers and memory content; 2) running the VeeR EL2 SoC platform in simulation mode, providing full visibility of all platform signals; and 3) the debug capabilities provided by the Vivado Design Suite, including the Integrated Logic Analyzer (iLA), which allowed tracing signals upon a predefined trigger event. The second hackathon workshop focused on educating participants in developing a memory-mapped accelerator integrated into the VeeR EL2 SoC platform using the Wishbone bus, as illustrated in ***Fig. 6***. The accelerator module was mapped into a designated address window and included a set of four memory-mapped registers: two data registers for passing integer data from the software to the accelerator, one result register for passing the result from the accelerator to the software, and a control register for activating the accelerator. The accelerator demonstrated a parallel vector addition operation, where each byte in the data registers and the result vectors represented an element in a 4-dimensional vector. As part of this workshop, the students were asked to extend the accelerator's functionality by 1) adding a vector multiplication operation and 2) extending each vector dimension (data registers and result registers) from 4 bytes to 8 bytes.

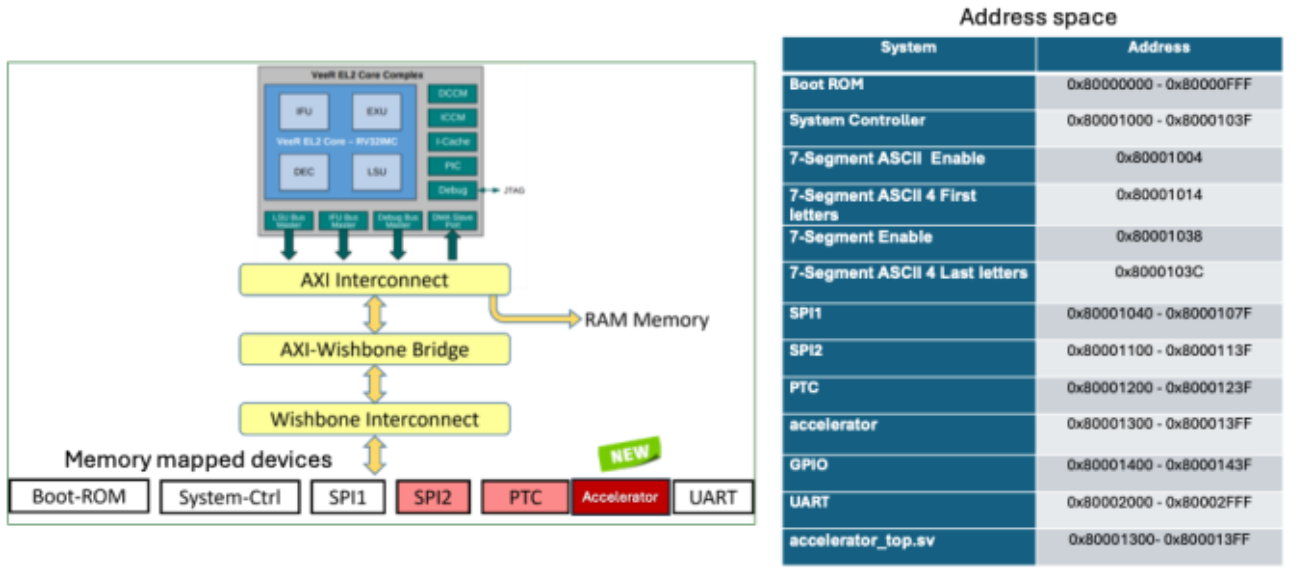

| System | Address |
| --- | --- |
| Boot ROM | 0x80000000 - 0x80000FFF |
| System Controller | 0x80001000 - 0x8000103F |
| 7-Segment ASCII Enable | 0x80001004 |
| 7-Segment ASCII 4 First letters | 0x80001014 |
| 7-Segment Enable | 0x80001038 |
| 7-Segment ASCII 4 Last letters | 0x8000103C |
| SPI1 | 0x80001040 - 0x8000107F |
| SPI2 | 0x80001100 - 0x8000113F |
| PTC | 0x80001200 - 0x8000123F |
| accelerator | 0x80001300 - 0x800013FF |
| GPIO | 0x80001400 - 0x8000143F |
| UART | 0x80002000 - 0x80002FFF |
| accelerator_top.sv | 0x80001300- 0x800013FF |

***Fig. 6*** Memory-mapped accelerator integrated into the VeeR EL2 SoC platform.

At the hackathon event, 30 students from the computer engineering and electrical engineering departments faced the challenge of accelerating the computation of the Fast Fourier Transform (FFT). The FFT was implemented in C code using fixed-point arithmetic on a series of 16 samples. Students were tasked with optimizing this implementation to achieve significant speedup. The winning criteria for the hackathon challenge were defined as achieving the best FFT speedup while minimizing area and power overhead. This required participants to balance performance improvements with efficient resource utilization, pushing them to apply innovative solutions and optimizations in their designs.

The participating students were required to complete a questionnaire consisting of open-ended questions, inviting them to reflect on their opinions regarding their hackathon experience, with an emphasis on the contribution of the hackathon to skill development. The overall feedback from the students indicates that the hackathon significantly contributed to enhancing their experience and deepening their knowledge of accelerator microarchitecture and design considerations. Additionally, the hackathon helped improve their understanding and experience with system-level design and software/hardware co-design, encompassing a programmable core and an accelerator.

5) Introduction to Computers and Computer Architecture at the University of Barcelona (UB)

The Computer Science degree at the UB includes the field of Computer Architecture. This field comprises three courses: Introduction to Computers (IC), Computers Architecture (CA), and Programming of Embedded Architectures (PAE). These three courses cover the essential knowledge that a graduate in Computer Engineering should have upon completing their studies. In the first two courses, RVfpga is widely used alongside the RISC-V Ripes simulator. In the third course, the ESP32-C3 is used instead of RVfpga, ESP32-C3 is a development board that integrates a 32-bit RISC-V microcontroller. Therefore, for this paper, we focus on the first two courses: Introduction to Computers and Computer Architecture, as both courses use RVfpga.

The Introduction to Computers (IC) is taken in the second semester of the first year. This course is preceded by two preliminary courses taken in students' first semester. The topics include:

- Basic Digital Design (BDD), where students learn the principles of combinational and sequential electronics, registers, buses, and ALUs. In BDD, students perform programming projects in Verilog, where they build state machines, culminating in the development of an ALU connected to a register bank and memory.
- Computer programming languages, primarily C and C++.

These two courses provide the foundational knowledge necessary to successfully complete IC. In the IC course, students learn about the different components of a processor and how they are connected, focusing on the RISC-V single-cycle model. Afterwards, the RISC-V ISA is studied using the Ripes simulator, which allows students to observe the actions of each instruction during each cycle. A brief introduction to compilers is included here, analyzing how they work. Thereafter, the control unit is studied, first for a RISC-V single-cycle processor and then for a RISC-V multicycle processor. The next block includes memory, types of memory, access methods, and memory hierarchy, concluding with peripherals, polling, and interrupts. Table *V* shows the different modules and their corresponding labs.

TABLE V
MODULES AND CORRESPONDING LABS IN THE INTRODUCTION TO COMPUTERS COURSE AT UB

|  |  |
|---|---|
| Module 1: CPU Structure | Lab 1. Introduction to Ripes Simulator |
| Module 2: RISC-V ISA | Lab 2. Exercises with Ripes; Introduction to RVfpga project |
| | Lab 3. Loops and advanced features; Introduction to compilers |
| | Lab 4. Subroutines; More about compilers |
| Module 3: Introduction to RISC-V microarchitectures | Lab 5. Same ISA, different microprocessors |
| Module 4: Memory, access methods, and memory hierarchy | Lab 6. Memory and Memory hierarchy |
| | Lab 7. Cache Memory; More about memory hierarchy |
| Module 5: IO Interfaces | Lab 8. Introduction to GPIOs |
| | Lab 9. Interrupts |

As can be seen in TABLE *V*, the modules covered in this course are accompanied by laboratory sessions. These lab sessions use a virtual Ubuntu machine equipped with the necessary tools: the Ripes simulator and the Visual Studio Code (VSCode) programming environment, which includes the PlatformIO plugin, which allows the use of the full functionality provided by RVfpga and the Nexys A7 development board.

Ripes offers a simple, user-friendly environment, greatly reducing the barriers to learning. Students can observe the execution of instructions and their effects on the different units: the register file, the ALU, the instruction and data memory, LIFO queue, and so on. This simulator aids in the understanding of the first two modules. Additionally, it helps students become familiar with the VSCode programming environment, as programs developed in Ripes can then be used on RVfpga and vice versa.

Another interesting exercise is the implementation of a program that calls a subroutine recursively, as is done in Labs 3 and 4. The goal is to study the behavior of the stack, memory accesses, and register usage. Subsequently, the same exercise is to be performed in C, and by using the Disassembly functionality, the 'Switch to assembly' option should be executed. This will allow students to compare the assembly code generated by the compiler with the original assembly code developed by them. This exercise helps to understand and explain the functionality of the compiler and its basic characteristics.

The microarchitectures' module is based on RISC-V and is built upon the knowledge acquired in both previous theoretical and practical sessions. It focuses on studying the evolution of execution time for specific assembly code executed in different microarchitectures [16]. This module is closely linked to Lab 5, where we use the Ripes simulator, which permits the use of different microarchitectures for the same code. This lab concludes the topics on microprocessors.

The module on memory and memory hierarchy explains the different types of memory, comparing access times, data hit rates, and memory accesses. We use for clarification the schema presented in the RVfpga Getting Started Guide, where the different memory types of the EH1 are explained. Cache memory access is introduced, detailing the various types of caches and explaining the operation of direct-mapped caches. Labs 6 and 7 are describe how to use the Ripes simulator for analyzing memory accesses. Students work with subroutines, using LIFO stacks pointed to by the stack pointer (sp) register, to save and restore registers in a subroutine. In Ripes, students can modify the cache size, mapping type (i.e., direct-mapped or associative), and the write policy (i.e., write-through or write back), to avoid inconsistencies between main memory and the cache. This module does not use RVfpga, because the instructors thought its use would be too complex for the students. The knowledge they acquire by the end of this module is enough to prepare them for using the various memory types available in RVfpga's VeeR EH1 microprocessor in the next course (CA).

Finally, module 5 covers input/output (I/O) interfaces by explaining them and the memory-mapped registers associated with them and describing how the microprocessor manages device handling. Polling, interrupts, and DMA are also covered. In this regard, theory is complemented by the RVfpga exercises presented in RVfpga Labs 1, 2, 3 and 6, where LEDs and 7-segment displays are used as outputs, and switches are used as inputs, detected through polling. Lab 9 analyzes interrupts using provided programs and exercises. Given the complexity of this last lab, a highly guided practical assignment is provided by the instructors.

The second course, Computer Architecture (CA), starts by using the concepts introduced in IC to explore them in depth. This course include 4 main modules, described in Table 2, that are covered in 26 theoretical lectures. Beginning with RISC-V-based microarchitectures, we examine methods to improve efficiency, such as pipelining, super-pipelining, and superscalar processors. We also explore the potential hazards these optimizations may introduce. We then cover the memory hierarchy, explaining how different types of cache memories work, exploring the various types of memory found in a computer, and concluding with memory management (SV32 and SV39). The course finishes with I/O interfaces & buses.

This CA course also includes 26 problem-solving lectures. In these sessions, in addition to solving problems from various textbooks, we introduce some of the material from RVfpga. Because the students do not have sufficient experience programming in hardware description languages (e.g., Verilog), these lectures are structured as master classes. The professor demonstrates part of the lab exercises to provide more examples and use cases for the students, then the students complete additional hands-on exercises.

RVfpga Labs 11 to 17 perfectly fit with module 1, processor optimization, as these labs cover the configuration, organization, and operation of the VeeR EH1 microprocessor, a 32-bit, 2-way superscalar, 9-stage pipelined in-order processor. However, for the master classes, we prefer to focus on the labs related to hazards, specifically labs 14 to 17.

TABLE VI
MODULES IN COMPUTER ARCHITECTURE AND LECTURES RELATED TO RVFPGA

|  |  |
|---|---|
| Module 1: Processor optimization | Labs 14 to 17 |
| Module 2: Memory Hierarchy | Lab 19 |
| Module 3: Memory management | Lab 20 |
| Module 4: IO Interfaces and Buses | Labs 6 and 7 |

RVfpga is also used in the rest of the modules. RVfpga Lab 19 fits into the Memory Hierarchy topic; it focuses on the instruction cache memory available in the processor. In addition, RVfpga Lab 20 addresses Memory Management, where two closely-coupled memories for instructions (ICCM) and data (DCCM) are tightly integrated with the core. As explained in the RVfpga documentation, these memories provide low-latency access and SECDED ECC (single-error correction and double-error detection error correcting codes) protection. Each of the memories can be configured as 4, 8, 16, 32, 48, 64, 128, 256, or 512 KiB. As with the previous module, these labs are defined and explained by the professor in a problem-solving lecture, where different memory access patterns are analyzed. Finally, the last module explains I/O interfaces and buses. RVfpga Labs 6 and 7 are used to improve the learning in this module. As in previous lectures, the Verilog programming section is conducted as a master class, where the instructor guides the students through the lecture.

### *B. Degree/Master projects*

In addition to the degree/master courses based on RVfpga, RVfpga is also an excellent resource to use in final projects related to computer architecture. We next summarize some projects that have been developed and successfully completed over the last few years, listed below and described in this section:

1. Master's thesis: FPGA implementation of an AD-HOC RISC-V system-on-chip for industrial IoT
2. Floating-point extensions for the SweRV EH1 core
3. Extending the functionality of the simulation tools for RVfpga
4. Half-Unit Biased format (HUB) format for floating-point for RVFPGA
5. Projects based on RVfpga at University of Sheffield
6. "Evaluation of Real Time Operating System in RISC-V", Instituto Superior Técnico, Technical University of Lisbon, Portugal (student: Rui Ferreira)
7. "Fault-Tolerant RISC-V Processor for Space Applications", Instituto Superior Técnico, Technical University of Lisbon, Portugal (student: Jaime Aguiar)

B.1 FPGA implementation of an AD-HOC RISC-V system-on-chip for industrial IoT.

In July 2020, Daniel León González, an IoT master's student at UCM, completed his master's thesis [18]. The project addresses the hardware implementation, using an Artix-7 FPGA, of a prototype IoT node based on RVfpga. It presents the implemented custom SoC and the development of the necessary Zephyr OS drivers to support a proof-of-concept application, which is deployed in a star network around a custom border router. End-to-end messages can be sent and received between the node and the ThingSpeak cloud platform. The thesis includes an analysis of the existing RISC-V processor implementations, a description of the required elements, and a guide to environment configuration and steps to build the complete project.

B.2 Floating-point extensions for the SweRV EH1 core

In July 2023, Alejandro Perea Rodríguez, a Computer Engineering student at UCM, completed his degree's thesis based on RVfpga [19]. The project is aimed at adding the FPnew floating point unit [20] in the RVfpga SoC. After completion of the processor extension, a series of verification tests were performed to check the correct operation of the new unit. The code developed in this project can be found at [21].

B.3 Extending the functionality of the simulation tools for RVfpga

In February 2024, Óscar Lobato Parra, a Computer Science student at UCM completed his degree's thesis based on RVfpga [22]. The project mainly extends the functionality of the RVfpga-ViDBo simulator by adding and verifying a tri-color LED. This device helps carry out in simulation the advanced practical exercises from Lab 8, where the PTC peripheral is used to generate a PWM signal that controls the RGB color shown by the tri-color LED.

B.4 HUB format for floating-point for RVfpga

In July 2024, Alfonso Martínez Conejo, a Computer Engineering student at the University of Malaga, completed his thesis based on RVfpga [23]. In this work, the floating-point addition and multiplication algorithms proposed in Lab 18 are adapted to the new HUB format through modifying the SweRV EH1 core. HUB is an emerging format that simplifies hardware and reduces execution time for implementations that require rounding to nearest. Its key feature is that rounding to nearest is achieved by truncation, eliminating the need for the hardware typically required for rounding in conventional formats (which involves a final addition with carry propagation). This reduction in hardware requirements results in shorter execution times, as demonstrated in this thesis.

5) Projects based on RVfpga at the University of Sheffield

We were keen to use the resources made available in this project in the Department of EEE at the University of Sheffield to upskill our students in HDL/SoC and design. However, we are not currently in a position to introduce a full module on this topic; given the existing requirements and breadth of the program (ranging from semiconductors to electrical machines), there is intense competition for the module slots that we teach and introducing new modules takes time and must meet a large need. So, instead we decided, initially, to offer final year undergraduate projects (BEng/MEng) and taught postgraduate (MSc) students extended projects based around the RVfpga resources. In all cases, the projects occupy one-third of an academic year's study. To facilitate the project work, we made the IP and RVfpga Lab resources available to the students (of whom there might be up 30 students in a cohort engaged on such projects). We provided small group support and opportunities for the whole group to meet together and to cover issues of interest to all students. We are keen for students to collaborate and to help upskill each other in the early stages of the project work when the objectives are to:

- Install and use the tools.
- Identify how to use (and specifically) adapt the IP for their own needs.
- Improve their use of HDLs.

Once the students move on to their own specific projects, the students work individually. More than one student may be working in the same basic area, so they must ensure that there is no hint of collusion on work that will contribute to their assessment. In this regard, the project supervisor is key in being able to establish (as the project progresses) that a student is responsible for their own work.

Projects are varied but often revolve around adapting the RVfpga SoC design to incorporate new functionality. This functionality may, at its most simple, be a new I/O component that sits in the memory space and is accessible from the RISC-V processor. In these projects, a student, initially, designs a component to implement a required function. An example from projects in 2023-24 is an ASCON lightweight cipher. The project steps were:

1. The core design was simulated (in Vivado) to establish functionality (***Fig.*** *7*).

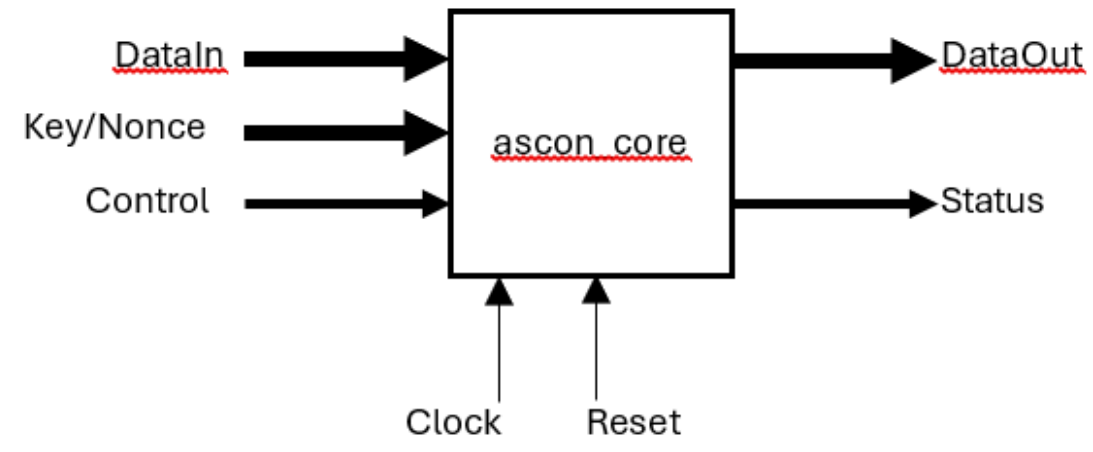

***Fig.*** *7* ASCON core.

2. Thereafter, it was wrapped in an appropriate memory mapped/register based interface allowing it to interface to the RISC-V processor (***Fig. 8***).

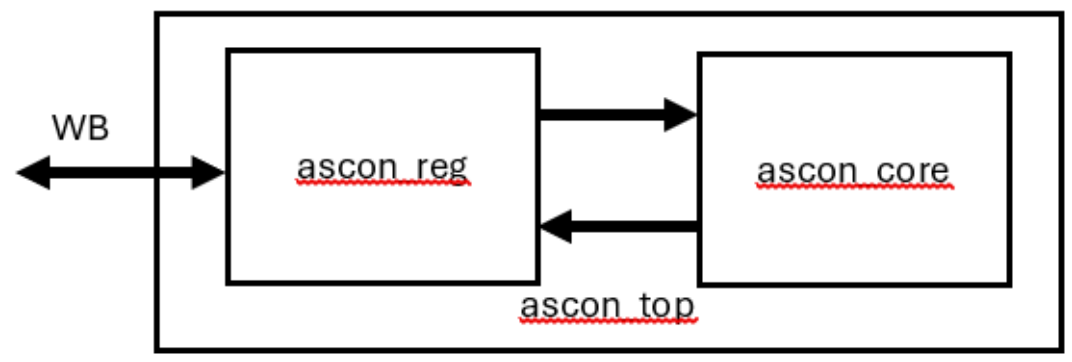

***Fig. 8*** Wrapped system.

3. The RVfpga design was then adapted (within Swervolf_core/wb_intercon) to implement the new component within the RVfpga SoC (***Fig. 9***).
4. This design was also simulated using Verilator, running C code that controlled the device.
5. Finally, the overall design was synthesized, implemented, and tested on an FPGA.

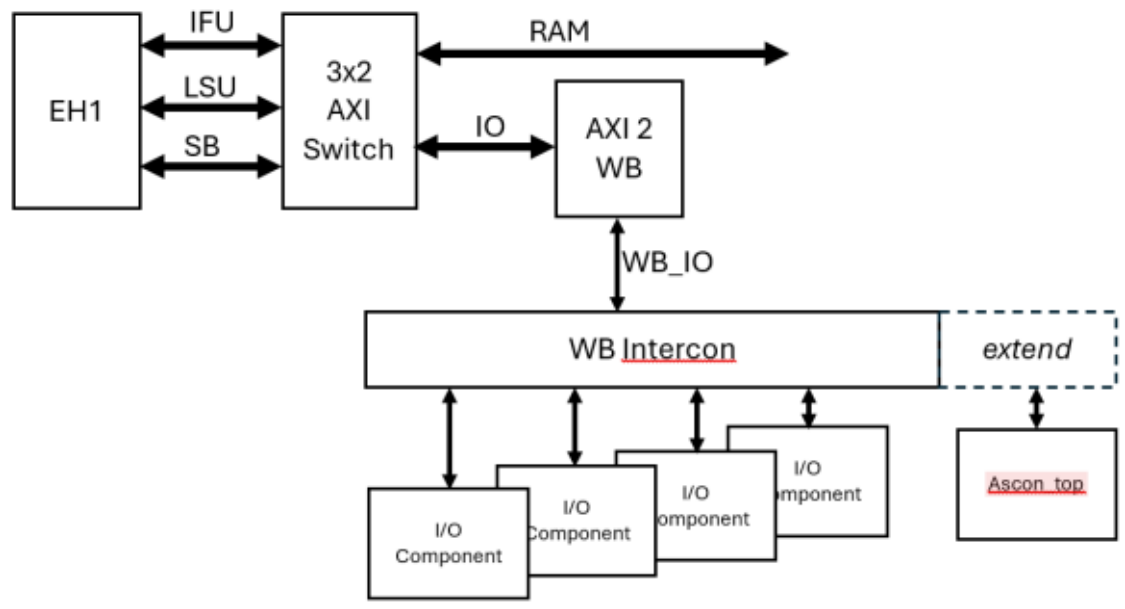

***Fig. 9*** New component of the RVfpga SoC.

As part of the design, students are asked to consider the following, in addition to the functional requirements: 1) the use case (how will the RISC-V SoC be expected to use the component efficiently and for what purposes); 2) testing (how will the design be tested and established to be functioning properly); and 3) performance.

Clearly, regardless of the high-quality documentation, the students who embark on these projects are climbing a steep learning curve. Although many of them have encountered Verilog in their previous courses, they are, for the most part, still novices. This situation requires significant support from their supervisors. Indeed, part of the motivation for introducing these projects has been to establish if this is a practical proposition or if, indeed, the learning curve is too steep. Our conclusion is that it can only really work effectively if the supervisor has become an expert, or can readily access someone who is. Moreover, we decided that projects must not complete the final step, 5: implementing the design on the FPGA. Depending on the design complexity and student capability, some projects never got beyond step 1: simulating the design. But, most typically, students got as far as step 4: simulating the design using Verilator.

A Case Study: One of the more complicated projects that has been undertaken by a masters student is a VGA display controller and simple graphics processor. This design (***Fig. 10***) is made more complicated by the requirement to implement two IP blocks and interface them to the rest of the system.

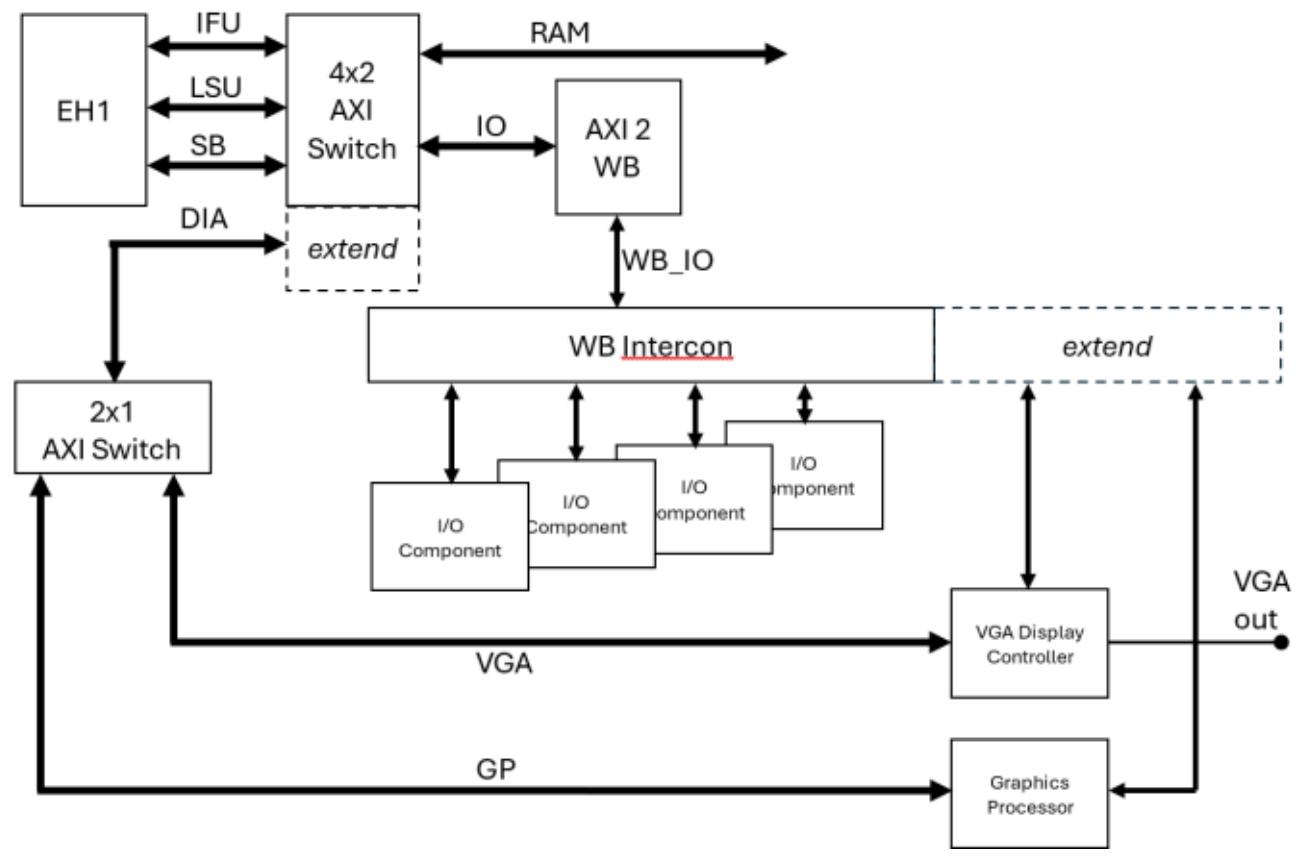

*Fig. 10* Overall design.

The two new memory-mapped components are:

- *VGA Display*: Collects data from a framestore in the system's RAM, generates timing, and displays pixels. The device exists as a set of registers that control resolution, pixel format, and timing parameters. Internally, a PLLE is used to generate the pixel frequency and a FIFO (which acts as a clock-domain crossing element between the system clock and the pixel clock).
- *Graphics Processor*: Passes commands for drawing lines and circles which cause the data in the framestore in the system's RAM to be updated.

Both of these components independently access the system RAM. To facilitate this, the AXI switch that combines/arbitrates if, ls, and debug memory accesses to RAM/IO space is extended to form a 4x2 switch. The spare slave port is connected to an additional 2x1 AXI switch that combines the master AXI port from the VGA controller (which is read-only) and the master AXI port from the graphics processor (which is read/write).

This project also required other changes to the system. For example, the axi2wb component (which is used in the simulation memory model) had to be extended to enable 64-bit writes.

Currently, this project is still a work in progress. It is running in simulation but has not yet been implemented on an FPGA.

6) "Evaluation of Real Time Operating System in RISC-V", Instituto Superior Técnico, Technical University of Lisbon, Portugal (student: Rui Ferreira)

System-on-Chip (SoC) implementations based on RISC-V processors can be extended with hardware accelerators to improve the performance of embedded systems by offloading heavy computational tasks from the processor. Using a Real-Time Operating System (RTOS) is essential in meeting the timing requirements and efficiently managing multiple tasks. In this thesis, the SweRVolf SoC was modified to incorporate a hardware accelerator for matrix multiplication. The modifications were mainly to the AXI-4 interconnect to provide an additional AXI-4 master port to control the hardware accelerator.

Initially, three scenarios were proposed for managing the hardware accelerator using Zephyr RTOS mechanisms that enable the RISC-V processor to execute different tasks, while the accelerator works in parallel. In the first scenario, a single thread accesses the accelerator and a semaphore is used to block the thread while it is working. The second scenario has several threads competing to access the accelerator and semaphores control access attempts. The third scenario uses a dedicated for managing the accelerator's access. Computational threads send requests to the manager thread to use the accelerator.

To evaluate these scenarios, two use case applications were developed and executed on the Nexys A7 FPGA board: matrix multiplication and a neural network to predict the digits of the MNIST dataset. The execution times of the different scenarios show that using a manager thread is more time-efficient than multiple threads competing for access to the accelerator, while providing a larger control over what thread uses the accelerator.

7) "Fault-Tolerant RISC-V Processor for Space Applications", Instituto Superior Técnico, Technical University of Lisbon, Portugal (student: Jaime Aguiar)

Recently, the viability of satellite launches and smaller space missions has increased, leading to a growing demand for more affordable onboard computer options. However, systems designed for space must be protected against radiation-induced data corruption.

This thesis explores the implementation of a Fault-Tolerant RISC-V system for space applications. The existing VeeR EL2-based RVfpga SoC by Imagination Technologies was thoroughly studied, and its implementation simplified. This effort was intended to reduce the EL2 system's complexity and facilitate modifications for doing research on fault-tolerant methods alternative to triple-modular redundancy. Various fault tolerance mechanisms, including one already implemented in the CPU, were examined through simulations and error injection via ICAP.

One of the main advantages of using VeeR EL2 is the fact that its Verilog code follows a modular structure in which even the CPU registers are instantiated as a component.

Even though the research was made on an oversimplified version of the VeeR EL2, it is possible to extend the fault-tolerance mechanisms to apply them to the original VeeR EL2.

### C. edX course

In year 2023, we released a MOOC based on RVfpga [11]. This MOOC was developed in edX and it includes 10 chapters that cover labs 1-4, 6-9, and 11 of the RVfpga course in Computer Architecture. The chapters provide instructions, hands-on tutorials, videos (including demonstrations), exercises, and multiple-choice questions.

The course begins in Chapter 1 with an overview of the course itself, the RVfpga system, and the tools, including detailed instructions with videos that show how to install and use the tools. The remaining ninc chapters (2-10) cover nine RVfpga labs, one per chapter: labs 1 to 4, about RISC-V assembly programming, and labs 6 to 9, about I/O, and Lab 11, an in-depth discussion of the VeeR EH1 core. These labs are called chapters 2-10 in the edX course. Each chapter includes several videos describing the principles and theoretical foundation of the topic covered in the chapter and showing demonstrations of how to use and practice these principles. Each chapter also includes multiple choice questions and exercises for practicing and gaining hands-on experience with the topics taught in the chapter.

Like the full RVfpga course, the edX course provides the entire source code (Verilog and SystemVerilog) for the RVfpga SoC. Course participants may use, modify, and execute this source code throughout the course, using the RVfpga simulation tools: RVfpga-ViDBo, RVfpga-Trace, RVfpga-Pipeline, and RVfpga-Whisper. Because the course may be completed in simulation only, the course can be completed without cost. Optionally, participants may also use the hardware, the Nexys A7 FPGA board, to test their modified RVfpga system and code that they generate throughout the chapters.

### *D. Microcredential based on RVfpga*

Microcredentials are small, targeted educational courses that allow users to certify that they have learned a specific skill or subject area. This kind of course is highly supported and recommended by the European Union [24].

In 2024-25, we will offer a new 6 ECTS microcredential at UCM called "RISC-V: Architecture and design based on VeeR commercial cores on FPGA", which has already been approved by UCM [25], and which includes 250 hours of training activities based on the RVfpga course in Computer Architecture and the RVfpga-SoC course in SoC design. Some important companies developing RISC-V processors have shown interest in the contents of this course and in hiring the students that will take it.

This microcredential contains 4 modules that follow a similar organization as RVfpga. The course begins with an extensive introduction to the RISC-V architecture and assembly programming, based on RVfpga labs 1-4, and continues with a series of input/output practices using the RVfpga SoC, based on RVfpga labs 5-10. Subsequently, a detailed study of the VeeR cores is carried out, using RVfpga labs 11-20. Finally, the deployment of a complete application is carried out, using a Zephyr Project as an RTOS as well as TensorFlow Lite for microcontrollers, based on the RVfpga-SoC labs.

### *E. Workshops, tutorials and webinars*

To increase accessibility to this RVfpga course, we have run several international "Train-the-Teacher" workshops and tutorials since May 2022, five in the U.S., six in Europe (the workshop from London was recorded and is available in [26]), one in Israel, two in Japan, one in Taiwan, and multiple events across China. With guidance from the authors, these hands-on events enabled attendees to quickly run and use the RISC-V system and its tools.

Although these are not the usual teaching activities, in the sense that they are not instructed in the context of undergraduate/masters courses, and these workshops are not aimed at students but instead at professors. These 1-day intensive practical workshops describe the main contents of the RVfpga course and guide students through subsets of select labs. Besides, these events have been a great method to obtain valuable feedback from experts in the field who are able to test the materials in place.

The workshop includes the most important aspects of RVfpga, describing the RVfpga system, showing attendees how to use and write code for it, and using hands-on practice to explore the peripherals and other features highlighted in the labs, including interrupts, performance counters, and benchmarking. Within the first hour of the workshop, users are able to write, compile, and run RISC-V programs on a commercial RISC-V core and SoC running in both hardware and simulation.

For the hands-on activities, we lend the attendees a Nexys A7 board (one board per student or per small group), where they can run the RVfpga SoC and complete all the labs. As for the RVfpga tools, we provide a Virtual Machine with everything already installed, that the attendees have to install and test the Virtual Machine on their laptops before the day of the workshop. In our experience after several workshops, this is the best approach, as it avoids many installation problems and saves a lot of time.

Responses from attendees include the following:

- "Excellent demonstration how real hardware and emulation matches, nice integration and IPC [instructions per cycle] and performance counter experiments."
- "I enjoyed the practical knowledge about RVfpga. My first successful FPGA project!"
- "A clear, focused, comprehensive [workshop] linking hardware, software, and RISC-V and relevant computer architecture within a workable toolflow and affordable hardware options."

In addition to the workshops, the authors hosted a Webinar sponsored by DigiKey to introduce RVfpga where 750 attended live, and more than 1,000 additional viewers subsequently watched the presentation [27].

## IV. RELATED WORK

In the latest years, due to the boom experienced by the RISC-V architecture, numerous educational resources have been developed to help users deepen their understanding of the RISC-V architecture and its applications. The vast amount of teaching resources currently available has clear advantages, as it allows users to access a multitude of options spanning various complexities and topics; however, it also presents disadvantages, such as the difficulty that users face in choosing the most suitable resource for their needs. Fortunately, RISC-V International seeks to coordinate efforts by centralize RISC-V teaching resources in a recently-launched repository [28], where a community-driven compilation of RISC-V resources and learning material is provided. The contents are continuously updated by the community and categorized based on the RISC-V scope and expertise level (from beginner to intermediate level), allowing anyone interested in RISC-V to discover the most relevant contents in an organized fashion. We should highlight that our edX RVfpga course is one of the recommended resources in the introduction ("After studying the Digital Design book in this section [beginner-level], you could jump to the intermediate-level edx RVfpga course if you wish as it expands on concepts discussed in the book."). In the same repository, a roadmap has been made available, which helps beginners to quickly select their own favorite materials for familiarizing themselves with RISC-V.

Several RISC-V learning materials, both theoretical and practical, are recommended in this repository [28]. As for the purely theoretical resources, the most relevant books on RISC-V computer architecture are included as a reference: Harris & Harris [29], Patterson & Waterman [30], and Patterson & Hennessy [31]. As for the practical resources, in addition to RVfpga (both the edX course [11] and the extended version [1] are referenced in the repository), other courses are also included.

The "Computer Architecture Basics" course [32], provides lectures, slides and videos about different topics such as arithmetic, CPU design, memory hierarchy, pipelined/speculative execution, and I/O systems, as well as nine tutorials and five labs based on the QtRvSim simulator. The tutorials describe, based on guided practical examples and short questions, the most important computer architecture concepts, such as RISC-V ISA and assembly language, memory hierarchy, pipelines and hazards, branch prediction, memory-mapped I/O, the PCI bus, and how to connect to an Avnet MicroZed board. Accompanying homework exercises delve into the concepts examined in the lectures and tutorials: number representation, image processing, processor pipelines, and code analysis. This is a useful and valuable resource, considered a beginner-level course, that explains many important computer architecture concepts, both theoretically and practically. It has some weaknesses, as it is mainly oriented to simulation, and peripheral support is very limited; it also uses an outdated bus (PCI) and does not address important architecture aspects such as superscalar execution, interrupt-based I/O, scratchpad memories, etc. In addition, although the lectures provide pdf/word documents, other materials are directly provided on the university's webpage (in fact, in some cases, the text is provided as an image), and some materials are protected and only available for students. Moreover, in some cases the course has not yet been ported to RISC-V and uses the MIPS architecture as it did in older versions.

Several resources included in the repository [28] show how to design and build a RISC-V processor of varying complexity and performance. One of those resources is learn-fpga [33], which provides two comprehensive courses (beginner-level: episode I and intermediate-level: episode II). The first episode starts from a very basic Verilog module that makes an LED blink and morphs it step by step into a basic yet fully functional RISC-V SoC implemented on an FPGA. The course also explains how to write programs in C and assembly for the RISC-V SoC. The second episode, explains how to transform the basic softcore from episode I into an efficient pipelined processor. As with the previous resource, these are very interesting and comprehensive materials with strengths but also some weaknesses. As opposed to the previous course, this one is based on an FPGA (the first episode can run on an affordable $40 Lattice Ice40HX1K whereas the second episode needs the larger Radiona ULX3S FPGA board) and can be also run in simulation (using Icarus Verilog or Verilator). Besides, it uses synthesis and place and route freeware tools (Yosys and nextpnr), a clear advantage for users. In terms of its weaknesses, the course is based on a non-professional and non-verified CPU (FemtoRV); it does not provide exercises but only guided activities; and some topics are not covered (such as interrupts, which are under construction, or advanced microarchitectural techniques such as superscalar execution).

Another option in the same line as learn-fpga is Quick Silicon's "RISC-V Processor Design" [34], a beginner-level course that provides skills and hands-on experience to the user to understand and design RISC-V-based processors from scratch. It provides several tutorials and videos that analyze fundamental concepts of computer architecture and processor design, the unprivileged RISC-V ISA, how to design the single-cycle RV32I-compliant YARP (Yet Another RISC-V Processor) core from scratch using SystemVerilog, and how to create RISC-V assembly programs and execute them on the designed processor. The course's weaknesses are a lack of important microarchitectural and SoC concepts, using a nonverified CPU, and perhaps most importantly, as opposed to most of the other resources in RISC-V International's repository, this course costs $60.

Another beginner-level option is [35], an edX course that shows how to create a RISC-V CPU with modern open-source circuit design tools (Makerchip online IDE), methodologies, and microarchitecture, all from a web browser. As with other edX courses, this course is free but can be optionally upgraded to be able to access graded assignments required for the certificate. The course does not use conventional HDL (such as Verilog or VHDL) but a specific language called Transaction-Level Verilog. It also relies on simulation only, and the CPU is simple and unverified. A similar course, which is not referenced in the repository, is [36], where the authors propose the creation of a pipelined RISC-V microarchitecture model using the same language and IDE as [35]: the TL-Verilog hardware description language and the Makerchip online development environment. The presented model said it would be taught during the 2023-2024 academic year, so at the time of writing this paper, results were not yet published. The course includes 11 labs about the following topics: RISC-V architecture and assembly language; design in TL-Verilog; RISC-V microarchitecture: pipelining, branches; memory access; and verification.

Finally, another beginner-level course also mentioned in the repository is [37] which, like previous resources, builds a general-purpose computer and uses it on a simulator and a CPU emulator; however, as opposed to previous resources, this course does not use RISC-V but a symbolic language called Hack.

An immediate-level resource provided in RISC-V International's respository is Shakti [38], an open-source initiative developed by the Indian Institute of Technology, which provides a complete development ecosystem where several processors, software tools, documentation and assignments are included. It provides an installation guide plus four assignments on RISC-V assembly programming, exceptions and interrupts, privilege levels, and paging and PMP. The main strength of this course is the variety of open-source processors provided. However, the course uses an Arty-7 100t FPGA board, which costs around $300, it only works in Linux (Ubuntu 18.04 is the verified distribution), and it does not cover many important computer architecture topics such as microarchitecture or memory hierarchy.

Finally, other interesting resources included in the repository, but which are orthogonal to the RVfpga courses, are related to RISC-V operating systems [39], RISC-V C compilers and toolchain [40] [41], and RISC-V extensions for vector applications [42] and for security [43].

RISC-V International, in association with the Linux Foundation, recently announced a certification [44] where the student is guided in the process of learning the RISC-V architecture and then complete an evaluation, called the RVFA exam, which demonstrates if the individual possesses the fundamental, entry-level knowledge and skills required of RISC-V hardware and software professionals. RISC-V International recommends several RISC-V Training Partners courses and RISC-V edX courses. Several courses, in addition to RVfpga, are included; however most of them require payment. Vicilogic [45] is an online RISC-V learning, assessment, prototyping, and course builder platform that uses remote FPGA hardware. Maven Silicon [46] provides a set of five RISC-V courses, three of which are free: "RISC-V ISA & RV32I RTL Architecture Design" that explores the architecture of RISC-V ISA and RV32I RTL design, "RISC-V RV32I RTL design using Verilog HDL" that explores RISC-V RV32I RTL design using Verilog HDL, and "RISC-V RV32I RTL Verification using UVM" that validates RISC-V RV32I RTL designs effectively with a UVM-based verification course. Finally, VSD-RISCV [47] offers courses that analyze the RISC-V architecture; the cost is $10 per course.

For the RISC-V edX courses recommended for the certificate [44], several resources are provided, most of which are free of charge. The three fundamental RISC-V courses [48] [49] [50] mainly analyze the RISC-V ISA and RISC-V assembly programming, as well as RISC-V tools, the GNU C compiler for RISC-V, and some basic concepts on OSs and libraries. In addition, some advanced edX courses are also listed, which analyze the microarchitecture of RISC-V CPUs [11] [35]. Embedded Linux, FreeRTOS, and Bare-Metal application development [51] [52] [53] [54] [55], and the RISC-V toolchain and compiler optimization techniques [56].

Finally, various RISC-V platforms have been introduced. These have not been included in the previous repositories, at least for the time being. In [57], the objective is to train students in hardware-software co-design of integrated systems. The course uses the Rocket Chip design platform, which allows students to begin with fast prototyping on a physical implementation, the Nexys A7 FPGA board. A demo is provided where students select an image and, optionally, a filter, using pushbuttons. The students may also select a filter and display and then view it on an VGA screen. Then, several labs are proposed on the following topics: getting started with the hardware and software environment; Exceptions, Traps and Interrupts; Multi-tasking, Multi-processing, Memory coherence and Caches; and adding Custom Hardware Peripherals. Finally, a project is proposed where the students are asked to develop the software to reproduce the behavior of the demo step by step: reading images from the SD card; image preprocessing and displaying; development of the Sobel filter and the convolutional neural network; and adding interrupts via the pushbuttons. These are interesting and comprehensive resources. The main problem is that the platform has not been updated for more than three years.

In [58] the authors provide an open-source RISC-V platform for education and research, which includes a configurable RV32IMC core and the MicroRV32 SoC, that, in addition to the core, includes several peripherals such as a UART, GPIO, an interrupt controller, and memory. They can be run in both simulation and on Lattice Semiconductor iCE40 FPGAs. The core has been verified via testbenches, software-based unit tests, example software applications, and formal verification, a cross-level approach in which the RTL and VP models are compared in a co-simulation setting. In addition to the sources of the core and SoC (which are coded in SpinalHDL, a Scala-based language for hardware description that extends the capabilities of traditional HDLs to provide less error-prone hardware design), several RISC-V assembly and C example programs are provided.

The BRISC-V toolbox [59] is the Boston RISC-V architecture design exploration suite. BRISC-V is comprised of several processor architectures, a simulator, and a visual Verilog file generation tool, for education and research projects. The Trireme Platform [60] includes everything you need to bring up the hardware and software of a custom RISC-V system, from ultra-low-power microcontrollers to high-performance multi-core processors.

Although RVfpga still has room for improvement, as we discuss in the next section, the RVfpga courses cover most of the gaps and content not included in these resources, such as advanced microarchitecture (superscalar execution, high-performance branch prediction, cache and scratchpad, benchmarking, interrupt-based I/O), support for low-cost FPGA boards (Basys 3 or Boolean), support for various simulation options (pipeline viewer, virtual FPGA board), use of commercial and verified cores (VeeR EH1 and EL2), etc.

## V. CONCLUSIONS AND FUTURE WORK

In this work, we have examined and discussed several teaching experiences using the RVfpga packages. We have also analyzed the latest computing curriculum guidelines developed by key organizations such as ACM and IEEE, and justified how the RVfpga courses align with these guidelines. Finally, we have conducted an in-depth examination of related work on RISC-V education and compared it with our own work.

As for our future work, we plan on making several improvements to the RVfpga course. In the short-term, we plan to release v3.1 that improves some parts of the current v3.0 RVfpga course in computer architecture. In the medium/long-term, we plan to release a v4.0, where additional extensions will be included, such as new labs about buses, exceptions, new peripherals (VGA, Ethernet…), and verification. We'd also like to extend the RVfpga-Pipeline simulator, by integrating the VeeR EH1/EL2 cores in the Ripes simulator, following the instructions from [61]. Finally, we also plan to review, improve, and extend the current RVfpga-SoC course.